# Head, posture, and full-body gestures in unscripted dyadic conversations in noise

Ľuboš Hládek[1,2], Bernhard U. Seeber[1]

*Audio Information Processing, Technical University of Munich, 80333 Munich, Germany*

*During writing at: Acoustics Research Institute, Austrian Academy of Science, 1010 Vienna, Austria*

*Corresponding Author: Ľuboš Hládek – lubos.hladek@tum.de*

# Abstract

Visual prosody may be critical for communication success in face-to-face conversations in noisy settings. Here, we explore the involvement of hand, head, and whole-body movements, as well as gesturing quality, in dyadic conversations in noisy settings. We hypothesize that increasing background noise would alter the frequency of conversation-related movements to support the roles of the speaker and the listener. Specifically, talkers may increase gesticulation and thus the use of hand, head, trunk, or leg movements more often, while listeners may increase backchanneling or head and trunk movements to improve the signal-to-noise ratio. Additionally, we test whether the synchrony between speech and hand gestures is affected by background noise. Here, pairs of normal hearing participants (n=8) stood in an audiovisual virtual environment while talking freely. The conversational movements were described using a newly developed labeling system with categories that respect their communicative function. The results showed higher gesturing rate during speaking than during listening. Increased levels of background noise led to increased hand-gesture complexity, modulation of head movements, and a change in trunk movements. People spoke 0.7 dB – 1.4 dB louder during hand gesturing in comparison to times with static drop posture but this was unrelated to presence of background noise. The analysis of hand-speech synchrony showed a modest decrease in synchrony for moderate noise level. People adapt their communicative behavior to increased background noise levels by increases in speech production levels and gesturing which may drive additional increase in speech production due to biomechanical coupling; listeners may increase backchanneling to support the exchange and their own signal-to-noise ratio. The synchrony analysis may reflect motivational factors of communication in noisy environments.

Keywords: gestures, body movement, dyadic conversation, cocktail party, hand-speech synchrony

# Introduction

Communication-related movements in face-to-face conversations play a central role in forming intentions and meanings. They may support exchange or signal misunderstanding. For instance, the widely used palm-up gestures may signal epistemic, presentational, or addressed meanings, but their exact meaning depends on the wider perspective(Clift, 2020; Cooperrider et al., 2018). In addition to gestures that carry explicit meaning, many natural and involuntary body movements, especially hand and head movements, synchronize with the speech. Fluctuations in speech acoustics that create prosodic contrasts often line up with instances of hand peak velocity (Pouw & Dixon, 2019). This type of gesture, called a beat or beating gesture, is characterized by swift movements and sudden stops. The beat gestures may signal stress in certain parts of speech and influence speech perception (Bosker & Peeters, 2021). Similarly, prosodic head movements may support rhythm, stress, and structure of the utterance (Munhall et al., 2004). Additionally, lip movements provide visual speech intelligibility benefit (Hadley et al., 2019, 2022; MacLeod & Summerfield, 1987). Hence, body movements support speech prosody by providing visual cues that may be more important in situations with degraded acoustics.

When communication becomes difficult due to acoustics, the movements and expressions respond to increased communicative demands. One of the well-studied mechanisms is the Lombard effect. Traditionally, it refers to vocal adaptations to increased background noise, hence degraded auditory feedback. The production level (Gardner, 1971; Hodgson et al., 2007; Weisser et al., 2021) and related aspects such as spectrum (Beechey et al., 2018; Webster & Klumpp, 1962) or timing (Hadley et al., 2019; Lu & Cooke, 2008) adapt but in conversational settings the production adapts also with increasing interpersonal distance (Gardner, 1971; Weisser et al., 2021; Weisser & Buchholz, 2019). This vocal adaptation thus also responds to the listener's demands, not only to the degraded self-feedback. This view has recently been extended to a multimodal perspective (Trujillo et al. 2021). Hand gesturing and facial expressions were modified by increasing background noise levels in a staged communication task. The hand gestures got more frequent and complex, while their data also showed that the speech production increased due to background noise only when people did not perform gestures, suggesting that increased gesturing compensated for increased voice production. Hence, the Lombard effect does not affect only vocal adaptation but also recruits the whole body and considers the listener's needs.

However, acoustic difficulty may influence communication through multiple mechanisms. Noise can degrade auditory feedback, increase cognitive load, or induce conversational repair. For example, under delayed auditory feedback, gestures show subtle timing shifts and reduced variability, consistent with dynamic entrainment between speech and movement systems, suggesting that gesture synchrony is sensitive to perturbations in auditory feedback. In Pouw & Dixon (2019), participants retold a story they had just watched (McNeill, 1992) while their speech and hand gestures were monitored with or without delayed auditory feedback. The speech-gesture synchrony, assessed by analyzing pitch and hand-velocity peaks, showed a delay and tightening of the coupling, attributed to increased entrainment. On the other hand, increased task complexity, for instance by reordering stimuli in a color-matching task, may produce higher desynchronization under cognitive load (De Jonge-Hoekstra et al., 2021). A cognitively demanding card game has also been shown to elicit spoken and gestural hesitations (Betz et al., 2023) which could be an aspect of desynchronization not observed before. These works indicate that speech–gesture coupling responds to production difficulty at the cognitive level. In face-to-face communication, adverse acoustic conditions can make conversation more effortful, but it is unclear whether similar mechanisms could be active in such situations.

From a biomechanical perspective, upper-limb movement and the respiratory–vocal system exhibit physical coupling. The myofascial-skeletal system is a dynamic system consisting of tensile and compressive elements and provides postural stability; therefore, it must be ready to react quickly to changes in equilibrium, such as limb extensions. Consequently, subglottal pressure, vocal intensity, and fundamental frequency can be affected via biomechanical connections (Pouw et al., 2020, 2021, 2025; Pouw & Fuchs, 2022). Further, vocal output may change even after alterations in posture, or in relation to leg movements (Pouw et al., 2021; Weston et al., 2024). These results imply that gestures and speech are also linked biomechanically, in addition to at the cognitive level.

Most studies rely on controlled tasks, designed utterances, or perception of phrases without context, which limit ecological validity. In face-to-face conversations, participants spontaneously adjust gesturing, head positioning, timing, and posture. They may naturally respond to changes in acoustic or communicative conditions, which is not possible in tightly controlled paradigms. However, empirical data on how people use various strategies in uncontrolled conversational settings and controlled acoustics are scarce.

The present study extends on prior findings by examining hand, head, trunk, and leg movements during unstructured dyadic conversations in noise. The analysis focuses on the perspectives of the speaker and the listener in face-to-face conversation in immersive virtual reality. First, we expect that as background noise increases, speakers adapt by increasing gesturing frequency and complexity (Trujillo et al., 2021) and may increase head movements by a similar principle. Also we expect listeners to recruit more visually relevant signals, such as backchanneling, by increasing nodding and recruiting trunk and head movements to improve SNR, such as by leaning forward or turning their head  Second, we test whether the speech production is affected by the biomechanical connections which would lead to increased speech production levels with increased gesturing regardless of background noise levels (Pouw et al., 2021) or whether the increased movements could compensate for the Lombard speech, when increased gesturing in noisy settings does not lead to increased speech production (Trujillo et al., 2021). Third, we test whether coupling of speech and gestures is strengthened as under delayed acoustic feedback conditions (Pouw & Dixon, 2019) or weakened as under increased cognitive load conditions (De Jonge-Hoekstra et al., 2021) and whether such change would be reflected in dynamic properties of hand gesturing such as peak velocity of gesturing in different noise conditions (Trujillo et al., 2021).

Hence, by exploring multimodal behavior covering multiple background noise levels, we aim to demonstrate how interlocutors strategically utilize gestures as a compensatory mechanism under adverse acoustic conditions and to examine the multimodal mechanism of the Lombard effect in a dyadic conversational setting.

# Methods

## Participants

Eight people participated in the experiment (four male, four female, 25.2±1.3 years (mean±SD), min 23 years, max 27 years). None of them had known hearing problems and their pure-tone hearing thresholds were tested in octave frequencies from 250 Hz - 8kHz using a clinical audiometer (MADSEN Astera², type 1066, Natus Medical Denmark Ap, Denmark). Each threshold was equal to or below 20 dB HL. The participants were recruited from university students and through personal contacts. During experiment, participants had conversation in English which was their second language. Each participant provided written informed consent, and the procedures were approved by the Ethical Committee of the Technical University of Munich (65/18S).

## Environment

The experiment took place inside an anechoic chamber (10 m x 6 m x 4 m; l x w x h) with the SOFE setup for virtual reality experiments with high-fidelity spatial audio. The audio system consisted of 61 loudspeakers mounted on a cube-shaped metal frame (4.39 m × 4.39 m x 3.4 m) – all pointing to the center. Thirty-six loudspeakers (Dynaudio BM6A mkII, Dynaudio, Skanderborg, Denmark) were distributed on a horizontal plane at approximate ear level (0° elevation), the remaining loudspeakers were distributed at elevations below and above the horizontal plane. The digital-to-analog converters (RME 32DA, Audio AG, Haimhausen, Germany) exciting the loudspeakers were controlled by a sound card (RME HDSPe, Audio AG, Haimhausen, Germany) in an external PC. The system further included four acoustically treated (32 dBA combined) video projectors (Barco F50 WQXGA, Barco, Kortrijk, Belgium) projecting on four acoustically transparent screens in front of the loudspeakers, creating a 360° projecting space with a floor area of 16 $m^2$. An optical motion tracking system (OptiTrack Prime 17W, NaturalPoint Inc. Corvallis, Oregon, USA) recorded the body movements (v3.01, Motive:Body, Optitrack, NaturalPoint Inc. Corvallis, Oregon, USA) of the participants at a 358.5 Hz sampling rate which was synchronized with the sound card’s word clock output (eSync 2, NaturalPoint Inc. Corvallis, Oregon, USA). A closed-circuit camera with an intercom was used to monitor the participants by the experimenter sitting outside in the control room.

## Audio-visual virtual environment

During the experiment, participants stood inside the SOFE projection area in audio-visual virtual space simulating the platform of an underground station (Hládek et al., 2021; Hladek & Seeber, 2022; van de Par et al., 2021). The acoustic simulation included reverberation of sources and their own speech (RT60 = 1.68, DRR = 2.7 dB, EDT = 0.31 s) and was created using the real-time Simulated Open Field Environment (rtSOFE) (Seeber et al., 2010; Seeber & Clapp, 2017; Seeber & Wang, 2021; T. Wang et al., 2022). The virtual environment was implemented in a game engine (Unreal Engine v5.2) with a custom visual model of the simulated underground station. The visual simulation was modelled according to the real underground station and included a running escalator (without escalator sound), train tracks, information boards, acoustic treatments on the side walls, an elevator, and other surroundings of the actual underground station. The model is freely available on Zenodo (Hladek & Seeber, 2022) .

The game engine provided the visual projection, while the acoustic simulation of the reverberated speech of both participants in rtSOFE received positional information from the game engine (Enghofer et al., 2021). The game engine controlled the simulation, and the plugin provided spatial updates for the acoustic simulation. The acoustic simulation was frequently updated according to the position of the participants (serving to update own reverberation profile) provided by the motion tracking system. The first order and higher order reflections up to 100 ms of the simulated impulse response were simulated with short latency using the image source method. Late reverberation (above 100 ms of the simulated impulse response) was obtained from a multi-channel recording of the environment (Hladek & Seeber, 2022). The details of acoustic modelling could be found elsewhere (Hládek et al., 2021). The speaker reverberation rendering further considered the directivity of human speech (Flanagan, 1960). The pick-up microphone of each speaker was spectrally equalized to the acoustic far-field for each participant prior to the experiment.

The acoustic and visual simulation was referenced to the middle of the simulation area which corresponded to the position R1 defined for the underground scene (Hladek & Seeber, 2022; van de Par et al., 2021).

## Experimental conditions

The experiment consisted of 30-minute-long conversations between two people standing on the platform of a simulated underground station. Broadband temporally modulated speech-shaped noise (Fastl, 1987) was played at different levels that changed every five minutes between these values: 0 dB SPL (No Noise, i.e. only the background noise of the video projection), 70 dB SPL (Medium), and 80 dB SPL (High). The order was pseudo randomized such that each condition took place twice. Reverberation added 2.3 dB; thus, the presentation levels of the noise conditions were 72.3 dB SPL and 82.3 dB SPL. The level of noise was calibrated by the loudspeaker calibration filters (from 100 Hz – 18 kHz) created with custom scripts, a measurement microphone (MM210, Microtech Gefell, Germany), and a microphone calibrator (Type 4231, Brüel & Kjær, Denmark). The levels were also verified with a portable sound level meter (XL2, NTi Audio, Schaan, Liechtenstein). Given the definitions of the underground scene (van de Par et al., 2021), the noise originated from azimuths 0°, 90°, 180°, -90° and simulated distance 1.6 m (note that the loudspeakers were physically at 2.1 m from the center). positions P1, P4, P7, P10 of the Underground scene 1, which corresponded to Each loudspeaker provided an independent sample of the noise that was generated from a custom-made script. The noise sources were reverberated and created by the convolution of noise samples with corresponding multi-channel impulse response of the given position in the underground scene.

## Procedures

Before the main experiment, participants received information about the experiment and signed informed consents. Subsequently, pure-tone thresholds of each participant were measured using pure-tone audiometry in a sound-proof booth. Following that, participants moved to the anechoic chamber building where they changed into the full-body motion tracking suits. The tracking markers and head-worn microphones were installed and calibrated by two experimenters. Participants wore head-set microphones (VT 800, Voice Technologies, Switzerland, on DWR-R02DN and DWT-B01N digital wireless transmitters, Sony, Japan) connected to the soundcard via an AD converter (Micstasy, RME, Germany). Additionally, the participants had full-body motion-tracking suits with the Optitrack Entertainment Marker Set installed on them (Core+Passive Fingers - 54 markers).

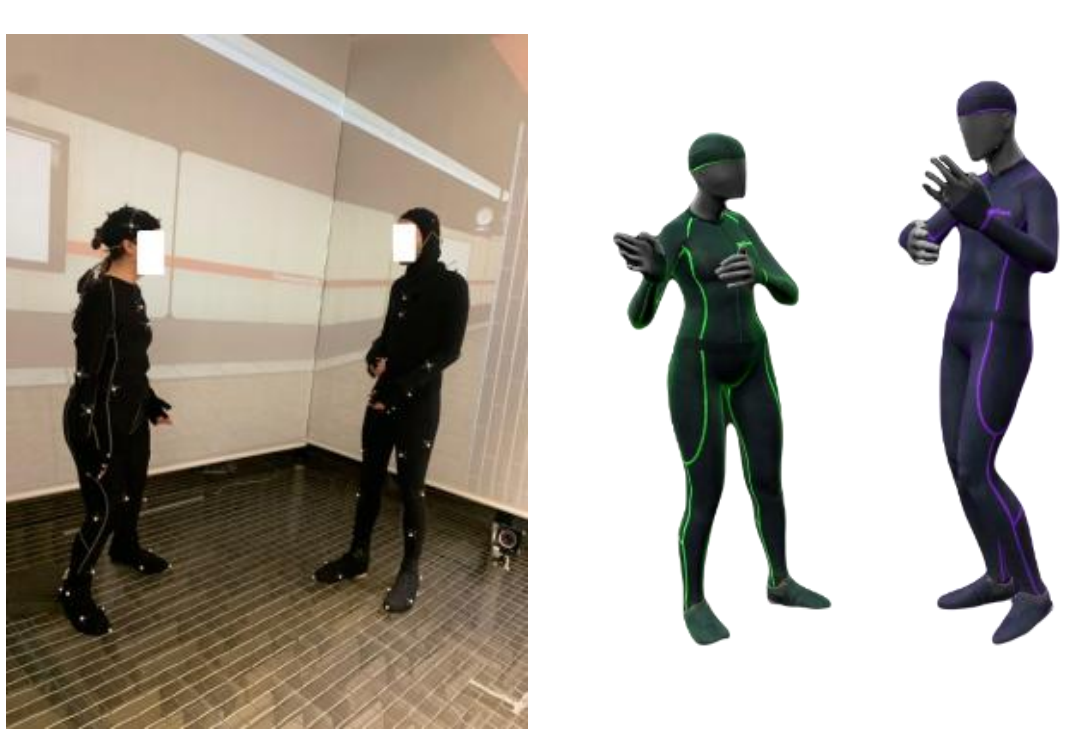

*Figure 1 A) Two participants are conversing in the virtual underground station, simulated using rtSOFE inside an anechoic chamber. B) Reconstruction of body postures (a different moment).*

The calibration of the marker set was done before the start of the experiment. The experimenters asked the participants to stand in the middle of the simulation area in a calibration pose. The correct positioning of the markers were tested by the pose reconstruction in the motion tracking software (v3.01, Motive:Body, Optitrack, NaturalPoint Inc. Corvallis, Oregon, USA). Next, the head-worn microphones were calibrated individually by asking each participant to stand one meter from a calibrated ½" measurement microphone (MM210, Microtech Gefell, Germany) and speak for 20 seconds while looking directly at the measurement microphone. Based on the power spectra difference between the reference and personal microphone, a finite impulse response filter with an order of 512 taps was designed, which was then applied in the acoustic simulation.

The main experimental task was to have a free dyadic conversation for thirty minutes. The task was to talk freely on any chosen topic but avoid topics that could be potentially harmful. For instance, they should talk about school and student life, or talk about the books they read, the places they visited, and movies they saw. They should avoid talking about politics, religion or any personal topics. After the calibration procedures which took place inside to audio visual space of SOFE , the experimenter left the room and initiated the experiment from neighboring control room.

The whole experiment was conducted in one session. The experimenters did not notice problems in the conversation flow, possibly because the participants had already spent the preparation time together in one room and were already acquainted with each other. Another factor that positively contributed to the conversation flow was that some participants already knew each other from school courses, but knowing each other was not a condition for participation nor an exclusion criterion in the experiment. Yet, most participants were strangers to each other before the experiment. After the experiment, the participants were debriefed regarding the purpose of the experiment.

## Gesture annotation system

We developed a novel annotation system to annotate typical movements during conversations with the focus on beat gestures (Elkjær et al., 2022; Hadley et al., 2019; Harrigan, 1985; Kendon, 2004; Lausberg & Sloetjes, 2016; Leonard & Cummins, 2011; McNeill, 1992; Wagner et al., 2014) . The motivation of development of such system instead of using an already established one e.g. NEUROGES (Lausberg & Sloetjes, 2009, 2016) was the focus on gesticulation (co-speech movement without explicit meaning) during conversation of two or three standing participants where all body parts including postures are affected (e.g., turning torso, crossed legs), including leaning forward and backward behaviors, and having simple categories that relate to speech communication.

Body motion was categorized in short time frames separately for the speaker and the listener. The speaking activity labels had only two options: whether the person was speaking or not and was determined using the detectSpeech function (v24.2, MATLAB, Natick, MA) while the output was resampled to the motion tracking sampling rate. The speech detection process was manually controlled, and misclassified periods of speaker silence due to cross-talk or unwanted noises were removed.

To be able to analyze hand, head, trunk, and leg movements movements during conversations, body motion analysis focused not only on head and hand movements, but on the entire body. Different body parts contribute distinctively to our expressions of intentions, states, and emotions, yet the combination creates a unique blend. Hence, we extended our scope from arm movements to head movements, trunk movements, and leg movements to capture the whole scale of movements and behavior in communication situations. This integrated framework enabled us to describe not only specific gestures but also broader postural changes, coordinated movements, and whole-body dynamics. The descriptions has been largely based on own observations during the piloting stage considering the previously developed approaches (Kendon, 2004; Lausberg & Sloetjes, 2009, 2016; McNeill, 1992).

### Arm movements and poses

The arm movements are a most prominent component of nonverbal communication (Wagner et al., 2014), conveying more information than speech. As arm contains seven degrees of freedom (Cheng et al., 2011; Z. Wang et al., 2018)—three at the shoulder, one at the elbow, one for forearm rotation, and two at the wrist—the range of possible movements is highly complex. In this study, we exclude finger movements and therefore consider a total of fourteen degrees of freedom across both arms. With the aim of describing typical movements in face-to-face communication situations, our analysis concentrates on the functional roles of hand and arm movements and uses their trajectories as a decisive cue. Based on this approach, we defined a structured set of gesture and posture categories that capture common expressive, emotional, and interactional movements of the body and the limbs considering other approaches in the literature (Kendon, 2004; Lausberg & Sloetjes, 2009, 2016; McNeill, 1992).

| Arm Movements | Description |
|---|---|
| Chopping/Clicking | Small, brief, and rhythmic hand movements with close relation to speech. They are often used to signal emphasis or highlight speech segments or express internal states or affection. |
| Circular Gesture | One or both hands swipe in a circular trajectory. Circular motions may illustrate continuity or the thinking process or accompany explanatory speech. |
| Complex Gesture | Both hands produce extended, multi-phase, or coordinated movements whose trajectories are not easily reduced to simple patterns defined by other |
| | gestures. These gestures often accompany expressive speech and can convey emphasis, uncertainty, affirmation, or support speech prosody – building expectations supporting the pace and speech rhythm. |
| Repetitive Gesture | Regular, repeated movement (often up–down or forward–back) of one or both hands, commonly serving as beat gestures that align with speech rhythm. In some contexts, repetitive motions may also express impatience or restlessness. |
| Palm-Inward Swipe | A hand movement where the palm faces inward (toward body) or downward. This gesture may function as expressive gesture and may accompany explanations and express intentions like assertiveness, regulation, or orders. |
| Palm-Upward Swipe | A hand movement similar to the palm-inward swipe, but performed with the palm facing upward (outward body). This gesture is widely associated with offering, requesting, questioning, or inviting further explanation. May be part of open posture. It may also mark openness, agreement, or acceptance. |
| Arm Postures | |
| Contractive Posture | A closed posture in which the hands rest in front of the body (e.g., arms crossed, hands clasped, or arms held close). The upper body appears more closed than in a neutral standing position. |
| Expansive Posture | An open posture with the arms extended outward, hands apart, or with the chest and torso widened. The body appears more open than when arms hang naturally. |
| Static Droop | A posture, usually standing still or ending a gesture to change into this posture, with hands hanging down naturally. |
| Self-Touching | Touching one's own body (e.g., chin, arms, thighs). Self-touch can signal cognitive processing, discomfort, anxiety, self-soothing, or mild itchiness. |

Table 1 Labels of typical arm movements in face-to-face communication and their description.

## Head movements

Head movements are a strong back-channel cue to keep the flow of communication. They often function for emphasizing, signaling attentiveness, and providing prosody. Biomechanically, head motion can be described through three rotational degrees of freedom—pitch, yaw, and roll—generated primarily by the rotation of the neck. Based on these fundamental movement axes, the present gesture labeling set is grounded in the trajectories and communicative functions of head movements, informed by established research on nonverbal behavior and movement kinematics (Kendon, 2004; Sharma et al., 2018).

| Head Movements | |
|---|---|
| Nodding | A common back-channel cue. An abrupt vertical movement and return of the head to signal reception and agreement. Additionally, nodding may express encouragement, or emphasis and can signal that the listener should continue speaking. |
| Shaking | A quick horizontal movement of the head. This gesture typically indicates disagreement (culturally specific); however, it can also express confusion or uncertainty. |
| Tilting | A rotation of the head around the roll axis to the left or right. Head tilts can convey interest, uncertainty, attentiveness, questioning, or emotional nuance, depending on context. |
| Turning | Rotation of the neck in the yaw axis, it can change the orientation of the face. |
| Up/Down | Rotation of the neck in the pitch axis to make the face towards up/down for a while. Different from nodding by motion velocity and communication function. |
| No Movement | The recorded person's head remains stationary or remains still without exhibiting any notable movements. |

Table 2 Labels of typical head movements in face-to-face communication and their descriptions.

### Trunk movement

Trunk movement involves coordinated motion of the waist, hips, and spine, allowing the upper body to rotate, bend, or lean. Trunk movement enables changes in posture, body orientation, and interpersonal distance during interaction, which is essential for increasing signal-to-noise ratio (Weisser & Buchholz, 2019) or head orientation benefit (Grange & Culling, 2016). The current set of trunk movements was selected according to their communicative functions. The movements may contribute to the expression of affiliation, social distancing, or simply may serve to improve speech intelligibility.

| Trunk Movements | |
|---|---|
| Turning Upper Body | A rotation of the trunk around the yaw axis, shifting the torso left or right and changing the orientation of the upper body. The movement may relate to other gestures (deictic) or may signal disengagement. |
| Leaning Forward | A forward tilt of the upper body is used to reduce the distance between the collocutors. The gesture may be associated with interest, affiliation, or serve to improve speech perception. |
| Leaning Backward | A backward tilt of the upper body that increases interpersonal distance. It may express discomfort, distancing, or simply adaptation of interpersonal space. |
| Leaning Side | A tilt of the upper body to the left or right. This may indicate comfort adjustments, or may signal shift in attention. |
| No Movement | A static posture with no noticeable upper-body movement during the observation period. |

Table 3 Labels of typical trunk movements in face-to-face communication and their description.

## Leg movement

Leg movement, like arm movement, can be described in terms of multiple degrees of freedom, with a single leg having seven rotational degrees of freedom across the hip, knee, and ankle joints. In everyday interaction, the communicative functions of leg movements tend to be more constrained than those of the arms. Here, we define a set of leg-movement labels based on the communication function. These behaviors contribute to posture, balance, and interpersonal signaling (Kendon, 2004).

| Leg Movements | |
|---|---|
| Standing On Tiptoe | Standing with the toes on the ground and the heels lifted. |
| Walking In Place | Minor stepping or lifting of the feet without shifting the body's overall position; often used to adjust posture or release restlessness. |
| Walking | Alternating leg movements that shift the body's center of mass horizontally, causing the person to change physical location. |
| Leg Raise | Lifting one foot off the ground while standing on the other leg. |
| Cross | Placing one foot adjacent to or crossing over the other (left crossing right or vice versa). |
| Full Turn | A rotation of the entire body by repositioning the feet and legs while keeping the upper body oriented until the rotation is complete. |
| No Movement | A static posture with no noticeable leg movements throughout the observed period. |

Table 4 Labels of typical legs movements in face-to-face communication and their descriptions.

## Gesture labeling

The motion tracking data served for gestural and temporal analysis. In the first step, we reconstructed virtual avatars from the data point clouds using the Optitrack software (v3.01, Motive:Body,

Optitrack, NaturalPoint Inc. Corvallis, Oregon, USA). Videos of the reconstructed avatars and aligned speech served for gesture annotation using annotation software (Lausberg & Sloetjes, 2009; Max Planck Institute for Psycholinguistics, The Language Archive., 2025). When recording the audio, we marked the motion tracking timestamp at the beginning of the recording, which allowed precise temporal alignment of the motion tracking data and the audio signal, hence the annotator listened to the conversation and watched reconstructed videos during the annotation.

The labeling was done by an annotator, who marked the beginning and ending of each gesture, including a pre-stroke hold, post-stroke hold, and retraction phase (Pouw & Dixon, 2019; Wagner et al., 2014). While labeling the four categories (arm gestures and postures were the same category), the annotator listened to the speech and watched the reconstructed video and was instructed to mark the whole phrase as a gesture - not only a stroke (Pouw & Dixon, 2019).

We analyzed the relative time of gesturing, which was determined as a proportion of time of the given gesture during one's own or the other collocutor's speech within a 10-minute interval (there were two repeats of 5 minutes in each condition). The values were transformed into a logit scale to account for the non-normal distribution of proportion values. Because of this transformation, we replaced the zero value (in case there were zero occurrences of a given gesture for a given participant in given condition) with a value of 0.005 (half a percent).

## Hand-speech synchrony analysis

We analyzed synchrony between the peaks of the pitch and the peaks of the speaker's hand movement velocity in beat gestures using a procedure motivated by the previous work (Pouw & Dixon, 2019) in a subset of arm gestures: *expansive gesture, complex gesture, circular gesture, palm-inward swipe gesture, palm-upward swipe gesture, chopping/clicking gesture, repetitive gesture*, since these represented primarily the beat moves. Gestures of one type that were spaced less than 50 ms were merged into one unit to correct for gaps in the annotations. The pitch traces were extracted using PRAAT (Boersma & Weenink, 2017) and temporally aligned with the motion tracking data. The pitches were extracted for the frequency range 51 Hz – 400 Hz and resampled to the motion tracking sampling frequency. The motion tracking data – hand position vectors – were filtered with a 10-Hz fourth-order Butterworth low-pass filter prior to computation of the first derivative.

The speech-hand synchrony was analyzed from the distribution of the hand-speech delay

$$d = t_{P_{max}} - t_{G_{max}} \; [ms] \quad \text{(Eq. 1)}$$

where $t_{P_{max}}$ corresponds to the time of pitch peak, which was a simple maximum of the F0 trace within the gesture duration, and $t_{G_{max}}$ corresponds to the time of the hand velocity maximum of the most prominent sub gesture. The most prominent sub gestures was found by finding local peaks in a two-second interval of hand velocity trace around the pitch peak (using MATLAB function *findpeaks*). More specifically, statistics $d$ was then defined as the signed time difference between the pitch peak and gesture stroke with the highest peak velocity.

$$d = \arg \min_{t \in [t_{P_{max}} - 1,\, t_{P_{max}} + 1] \cap [dur_G]} \left| t - t_{P_{max}} \right| \quad \text{(Eq. 2)}$$

such that

$$v_G(t) = \max_{t' \in [t_{P_{max}} - 1,\, t_{P_{max}} + 1] \cap [dur_G]} \left( \left| \frac{dG(t')}{dt'} \right| \right) \quad \text{(Eq. 3)}$$

Here $G$ corresponds to the gesture trace and $v_G(t)$ is the gesture velocity. The two-second interval was chosen based on the distribution ranges observed by (Pouw & Dixon, 2019). We assumed that delays above or below 1 s do not represent a synchronous relationship.

# Results

## Speaker’s and listener’s gesture frequency

The relative frequency of the individual gestures was analyzed across the three conditions as a proportion of the time of the given gesture relative to the time when the gesticulator was speaking (Gestures As Speaker) and relative to the time when the co-locutor was speaking (As Listener).

### Hand gestures

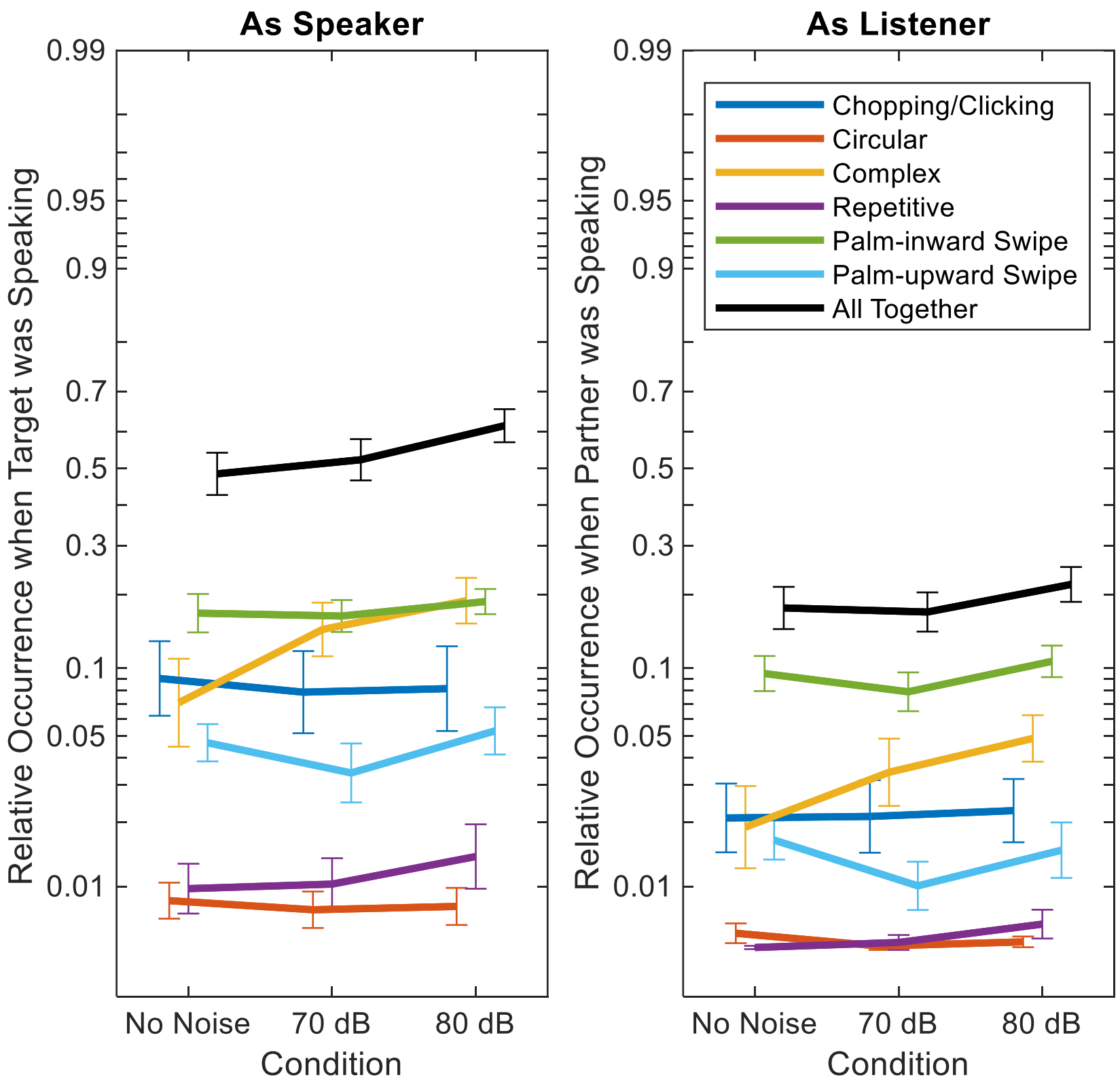

*Figure 2 Hand gesture usage as a function of increasing background noise level (x-axis). The y-axis corresponds to the relative amount of time people were gesturing during speaking (Left: As Speaker) or when the conversation partner was speaking (Right: As Listener). Color codes individual gesture types (see legend), the black line corresponds to the relative time when either of the selected gestures (in the legend) were performed. The proportional data were Logit transformed before plotting, and the y-axis was then rescaled to show back-transformed proportions in this and the following graphs. Error bars represent the standard-error of the mean (here and following graphs).*

Figure 2 shows the usage of different hand gestures under varying noise levels of background noise in two modes: either during speaking (Left) or listening (Right). The data show proportions of time for each gesture that represented beat gestures. The data were transformed into Logit scale before plotting and for statistical analysis, the y-axis shows proportions (e.g., 0.9 corresponds to 90%).

Statistically significant differences were found between the hand gesture types (Table 5: main effect of Type) modulated by the background noise (Table 5: interaction Condition x Type). Palm-inward swipes and complex gestures were more frequent than palm-upward swipes (pair comparisons using t-test results in $p<0.05$ after Bonferroni correction), and palm-inward swipes were more frequent than chopping/clicking gestures ($p<0.05$ after Bonferroni correction) while repetitive and circular gestures (significantly different from all other gestures at $p<0.05$ after Bonferroni correction) were rarely present. Complex gestures increased with the background noise, while the other gestures were affected to only a small extent.

As expected, people gestured significantly more during speaking than listening (Table 5: main effect of Mode), and this was evident for all observed types of gestures, however, the increase during speaking relative to listening was more pronounced for the complex gestures, than for instance, circular gestures.

| Source | F value | p value | p-adjusted | Significance | Partial $\omega^2$ |
|---|---|---|---|---|---|
| **Mode** | 123.96 | 0.0000 | 0.0000 | *** | 0.9461 |
| **Type** | 32.61 | 0.0000 | 0.0000 | *** | 0.8187 |
| **Mode x Type** | 10.86 | 0.0000 | 0.0003 | *** | 0.5848 |
| **Condition x Type** | 3.02 | 0.0031 | 0.0248 | * | 0.2239 |

Significance levels: *** $p<0.001$ * $p<0.05$

Table 5 ANOVA of the relative occurrence of hand gestures.

## Hand postures

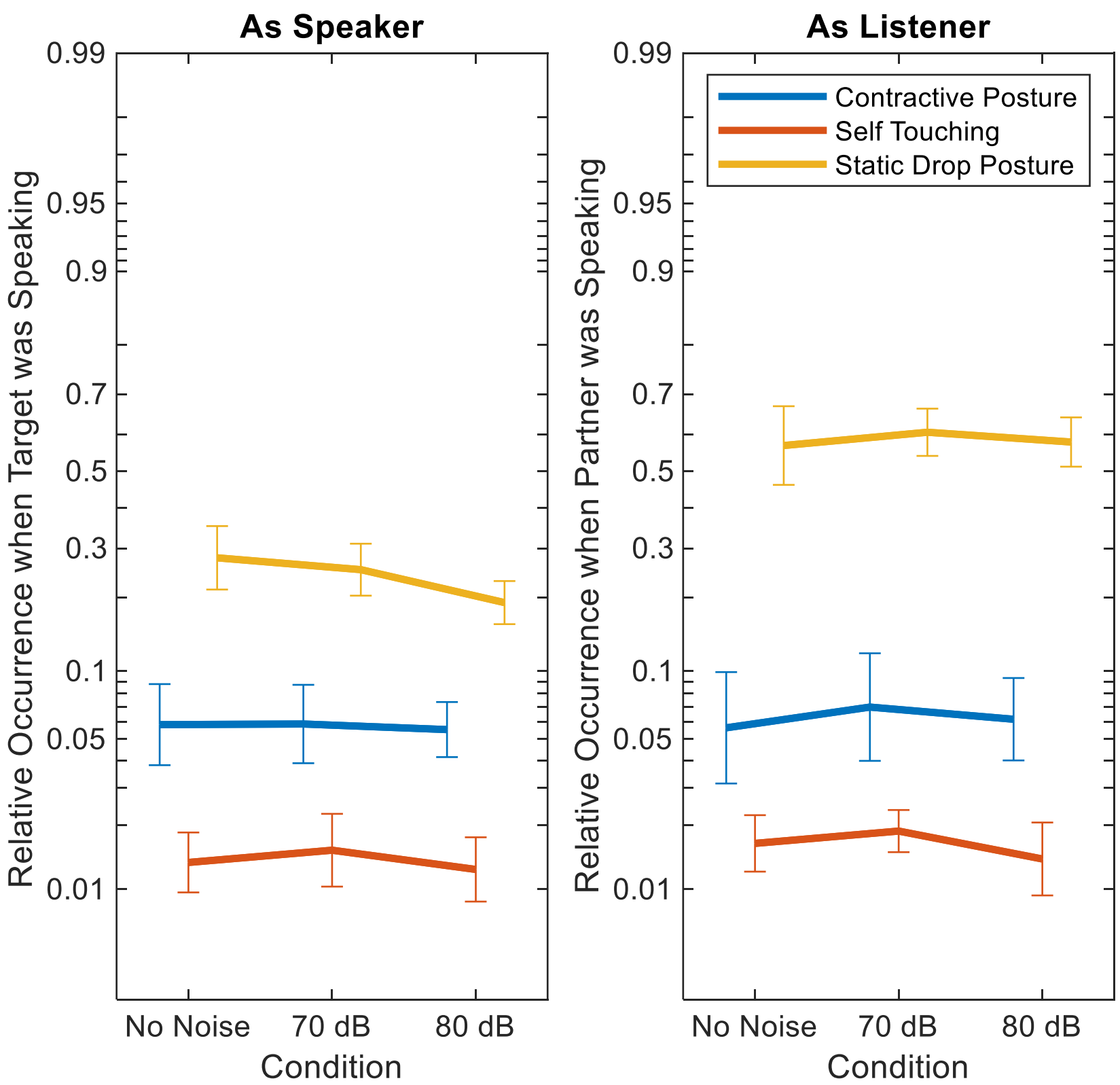

*Figure 3 Hand postures as a function of increasing noise level (x-axis).*

Figure 3 analyzes hand postures (which were in the category of hand movements) during speaking (left) and listening (right). Color codes posture type (see legend), x-axis shows condition.

The static drop posture was the most common occurring more often than contractive posture or self-touching (Table 6: main effect of Type; pairwise post-hoc comparison using t-test showed statistically significant differences at level $p<0.0.5$ after Bonferroni correction). Its frequency of occurrence decreased during speaking compared to listening (Table 6: interaction Mode x Type) complementing the increase in gesturing during speaking.

| Source | F value | p value | p-adjusted | Significance | Partial $\omega^2$ |
|---|---|---|---|---|---|
| **Mode** | 16.95 | 0.0045 | 0.0045 | ** | 0.6950 |
| **Type** | 32.04 | 0.0000 | 0.0000 | *** | 0.8160 |
| **Mode x Type** | 14.77 | 0.0004 | 0.0021 | ** | 0.6630 |

Significance levels: ** <0.01 *** <0.001

Table 6 ANOVA of the relative occurrence of hand postures.

## Head gestures

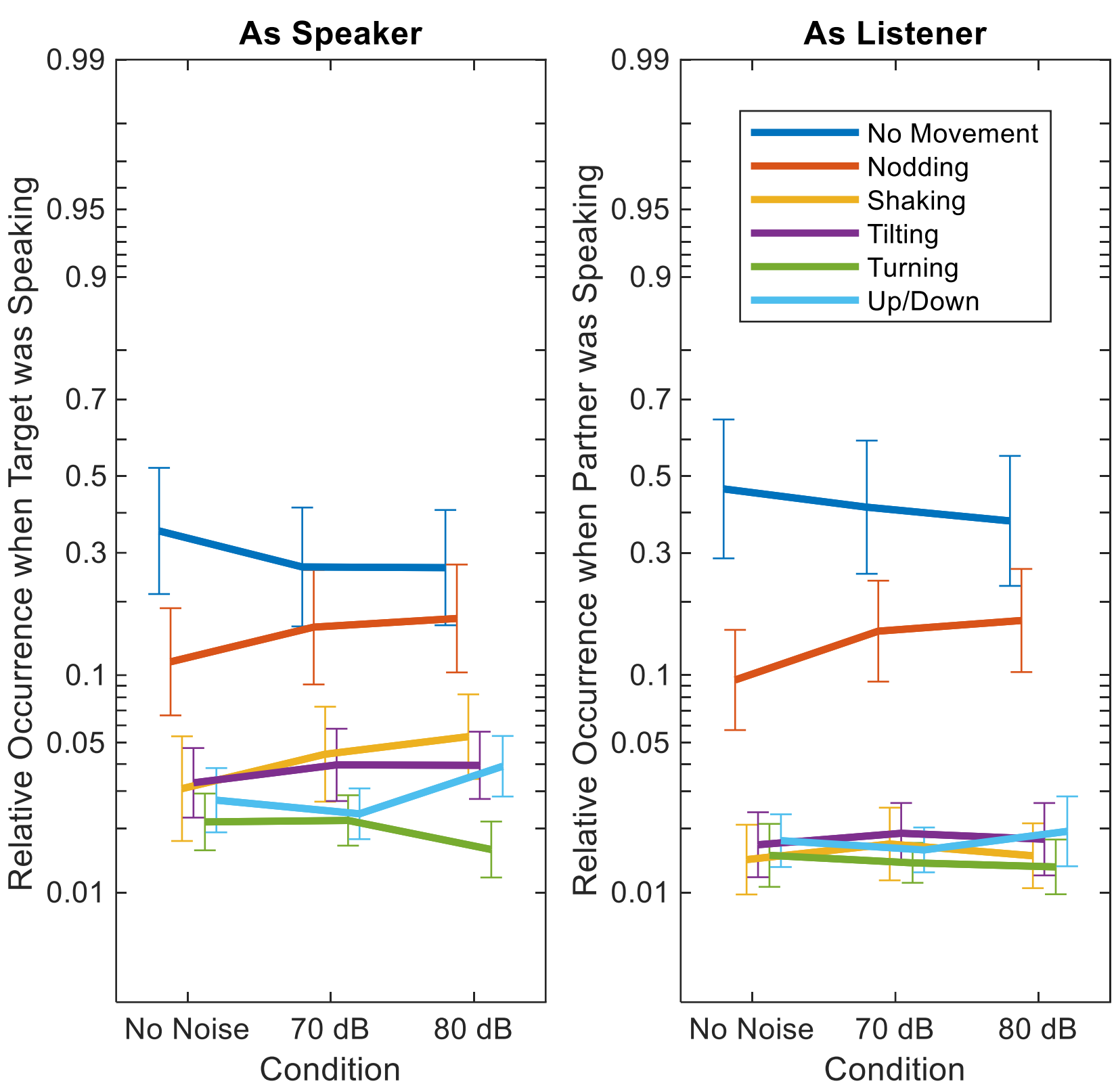

*Figure 4 Head gestures as a function of increasing noise level (x-axis). The y-axis corresponds to the relative amount of time people performed the posture during speaking (Left- As Speaker) or when the conversation partner was speaking (As Listener). Color codes posture types (see legend).*

Figure 4 shows frequency analysis head gestural movements during speaking (left) and listening (right). Nodding was the most frequent head movement. Other gesture types (shaking, tilting, turning, up/down movements) were rarely present during listening, but more often present during speaking, as evidenced by the statistically significant interaction of Mode and Type (Table 7), while both main effects were also significant. Head gesturing type was also modulated by the presence of the background noise (Table 7: Condition x Type interaction). The interaction was driven by the increase of turning complemented by a decrease in up/down motion at the high background noise level.

| Source | F value | p value | p-adjusted | Significance | Partial ω² |
|---|---|---|---|---|---|
| **Mode** | 19.94 | 0.0029 | 0.0029 | ** | 0.7301 |
| **Type** | 21.13 | 0.0000 | 0.0000 | *** | 0.7420 |
| **Mode x Type** | 9.36 | 0.0000 | 0.0011 | ** | 0.5442 |
| **Condition x Type** | 4.00 | 0.0002 | 0.0091 | ** | 0.2997 |

Significance levels: ** <0.01, *** <0.001

Table 7 ANOVA of the relative occurrence of head gestures.

## Trunk postures

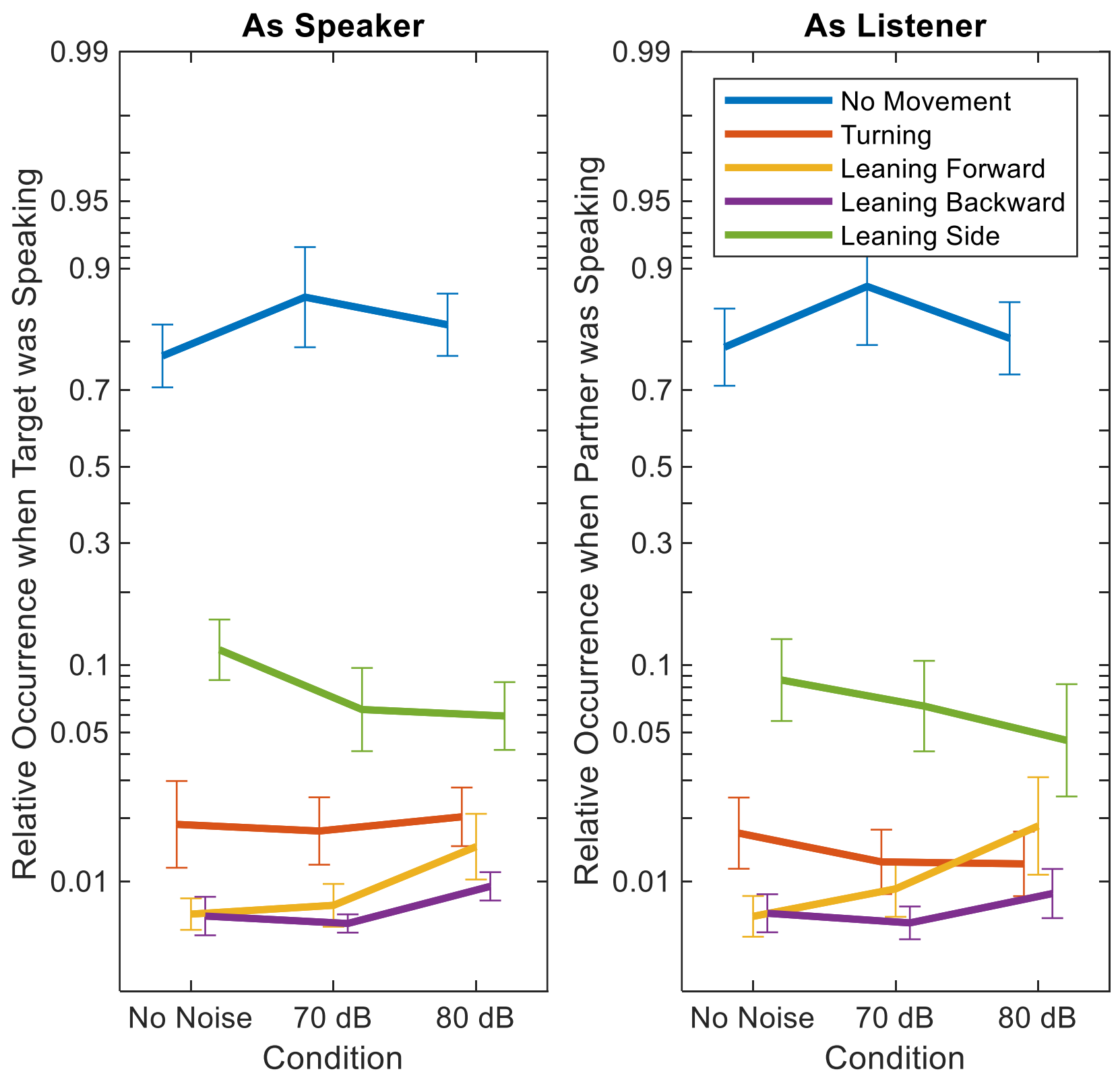

*Figure 5 Trunk movements for different posture types (see legend) as a function of increasing noise level (x-axis).*

Figure 5 shows the analysis of trunk movements during speaking (left) or listening (right). Leaning side was the most frequent trunk motion during conversation, significantly more than leaning forward and leaning backward (Table 8: main effect of Type, the mentioned pairwise comparisons were statistically significant at $p<0.05$ after Bonferroni correction). We saw a statistically significant increase the use of leaning behavior with the increased noise level (Table 8: Condition x Type). The interaction is not significant after the Geisser-Greenhouse correction but the lack of significance after correction could be also related to overall low incidence of this type of movements in the dataset hence, we interpret it as an important finding.

| Source | F value | p value | p-adjusted | Significance | Partial $\omega^2$ |
|---|---|---|---|---|---|
| **Type** | 61.04 | 0.0000 | 0.0000 | *** | 0.8956 |
| **Condition x Type** | 2.24 | 0.0374 | 0.1239 | n.s. | 0.1508 |

Significance levels: *** <0.001

Table 8 ANOVA of the relative occurrence of trunk movements.

## Legs movements

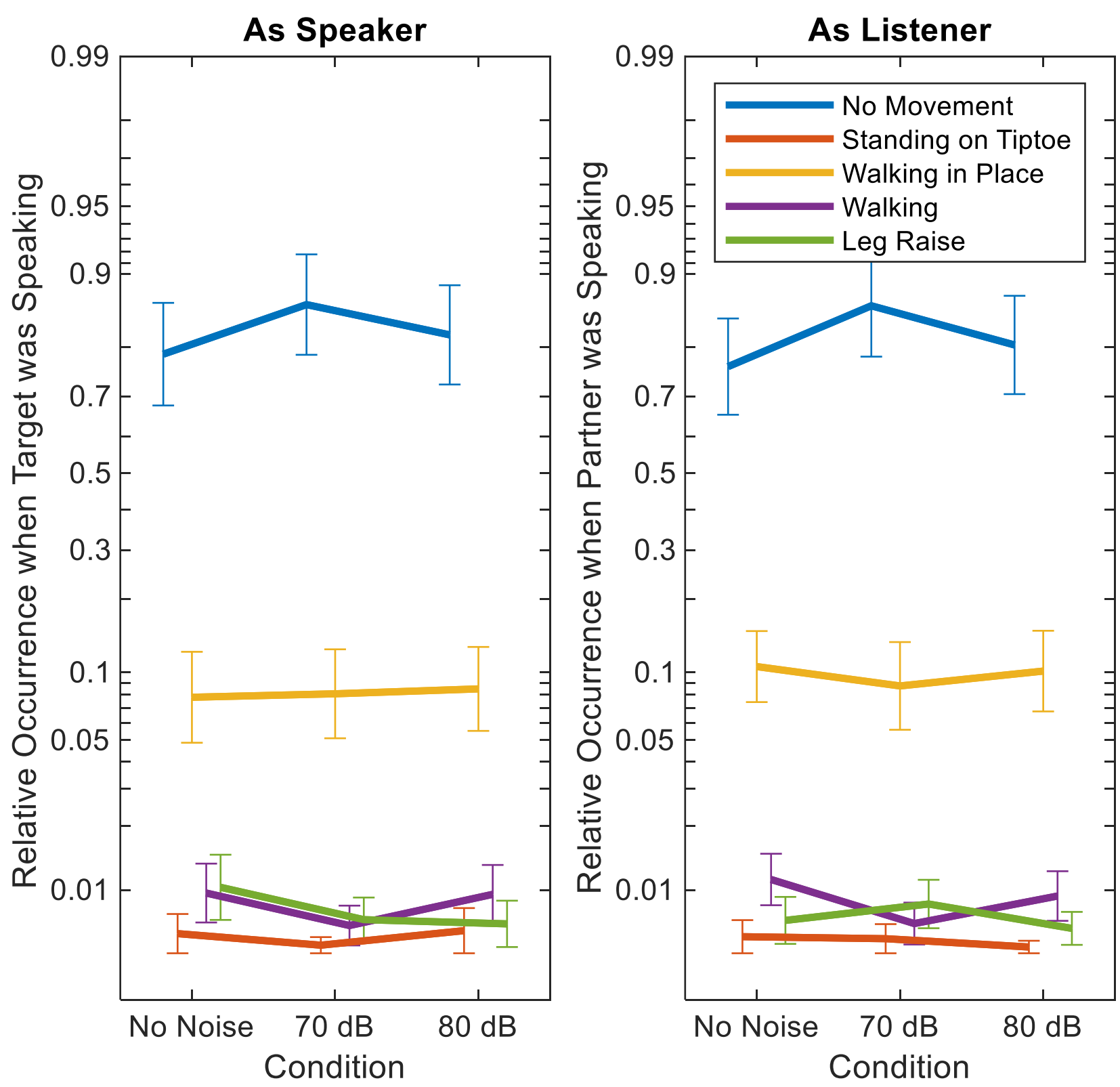

*Figure 6 Legs movements (see legend for type) as a function of increasing noise level (x-axis).*

Figure 6 shows the analysis of leg movements as a function of the background noise level (x-axis) and mode (As Speaker – left, As Listener - right). Walking in Place was the most common type of movement while the other types were rare (usually less than 1%) (Table 9: main effect of Type). Neither the mode or the background sound level significantly affected leg movement frequency. Although the Condition and Type interaction was close to significance, it may also relate to zero incidence of some of the leg movements in some of participants that would artificially decrease variance and additionally the data do not show any clear trend with respect to our conditions.

| Source | F value | p value | p-adjusted | Significance | Partial ω² |
|---|---|---|---|---|---|
| **Type** | 61.02 | 0.0000 | 0.0000 | *** | 0.8956 |
| **Condition x Type** | 2.05 | 0.0570 | 0.1379 | n.s. | 0.1302 |

Significance levels: *** <0.001

Table 9 ANOVA of the relative occurrence of legs movements.

Speech production by gesture type

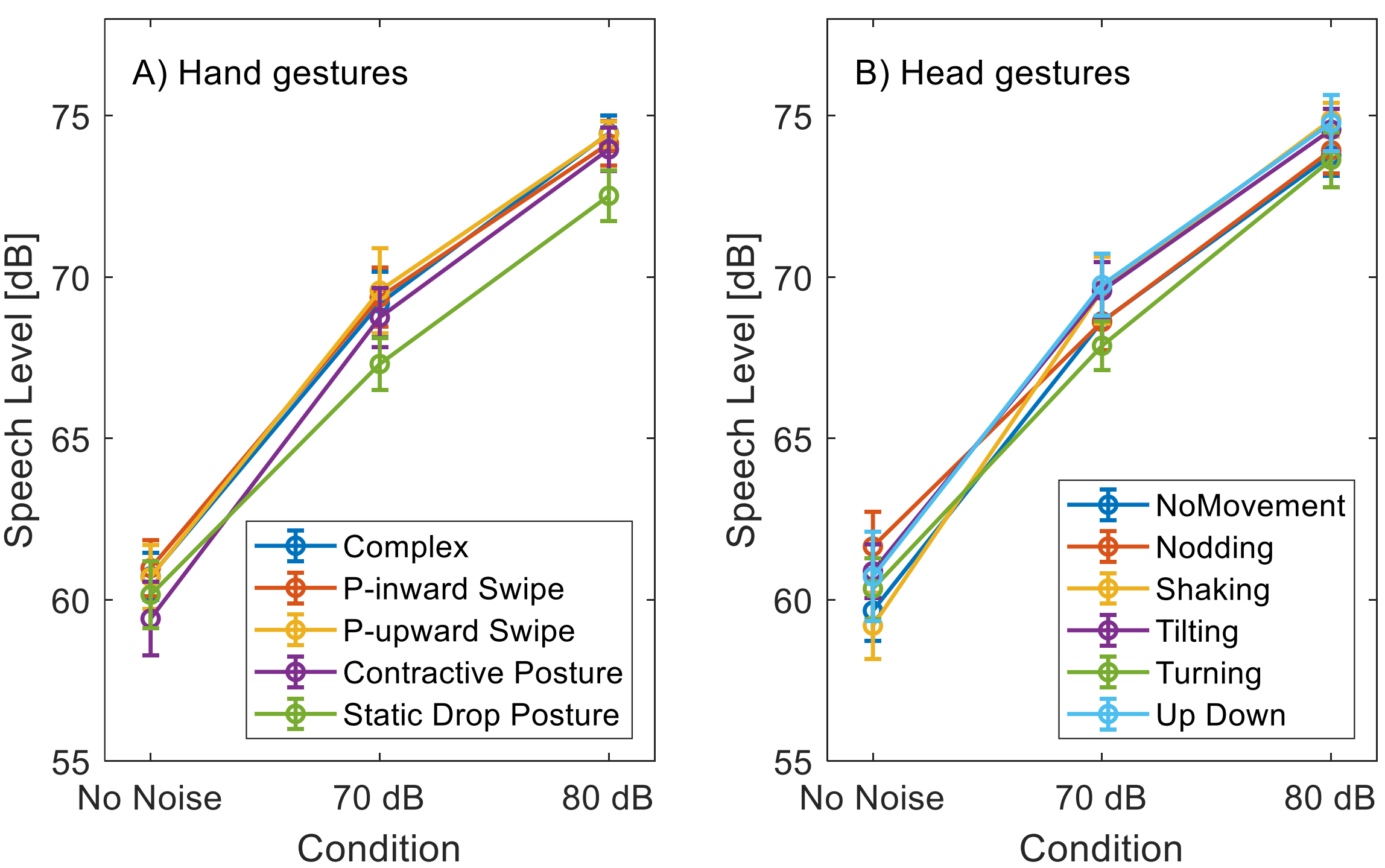

*Figure 7 Speech production change as function of background noise level. (A) Line colors code different hand gestures or hand postures. (B) Line colors code different head movements.*

Figure 6 shows speech production levels as a function of background noise level and gesture type. The aim of the analysis is to assess whether speech production increases with background noise level as is expected from degraded self-feedback in Lombard effect and whether active gesturing further increases production levels as expected from biomechanical coupling regardless of the effects background noise. The speech production could be further decreased by compensatory multi-modal effects. Some gesture types, as well as trunk and leges movements, were not present in all conditions for all participants and therefore were not considered for this analysis.

The data show that a statistically significant effect of background noise (Table 10 – main effect of condition), we observed an increase o 8 dB between No Noise and 70 dB condition and further 5 dB increase to 80 dB condition. The data further show a statistically significant main effect of gesture type (Table 10 – main effect of type). More specifically Complex Gesture and Palm Swipes were on average 68.1 dB and were not statistically significantly different from each other, but the postures without movement Contractive Posture (67.37 dB) and Statis Drop posture (66.7 dB) were offset from the other three hand-movement-related gestures and were also statistically different from all other types (pairwise post-hoc comparisons were done with a Bonferroni correction, $p<0.05$). Hence, we see an increase from 0.7 dB to 1.4 dB related to hand movements but no modulation by the background noise condition of these differences. The ANOVA showed a statistical difference but only without correction of violation of sphericity. The latter tells that if the effect the effect size is smaller than what could we capture with current sample size.

| Source | F value | p value | p-adjusted | Significance | Partial $\omega^2$ |
|---|---|---|---|---|---|
| **Type** | 8.20 | 0.0002 | 0.0056 | *** | 0.5070 |
| **Condition** | 205.96 | 0.0000 | 0.0000 | *** | 0.9670 |
| **Condition x Type** | 2.42 | 0.0253 | 0.0977 | n.s. | 0.1688 |

Significance levels: *** <0.001

Table 10 ANOVA of the speech production related to hand gestures in different noise conditions.

The analysis of speech production in relation to head movements also showed a strong effect of background noise level (Table 11 – main effect of condition). Similarly to the above analysis of speech production with respect to hand gestures, the increase reflects the classical Lombard effect. Again, we see indications of slight effects of type and the interaction of condition and type but after considering the statistical adjustments these effects were not significant.

| Source | F value | p value | p-adjusted | Significance | Partial $\omega^2$ |
|---|---|---|---|---|---|
| **Type** | 2.49 | 0.0494 | 0.081 | n.s | 0.1759 |
| **Condition** | 190.53 | 0.0000 | 0.0000 | *** | 0.9644 |
| **Condition x Type** | 2.76 | 0.0062 | 0.0612 | n.s. | 0.2012 |

Significance levels: *** <0.001

Table 11 ANOVA of the speech production related to head movements in different noise conditions.

## Hand-speech synchrony

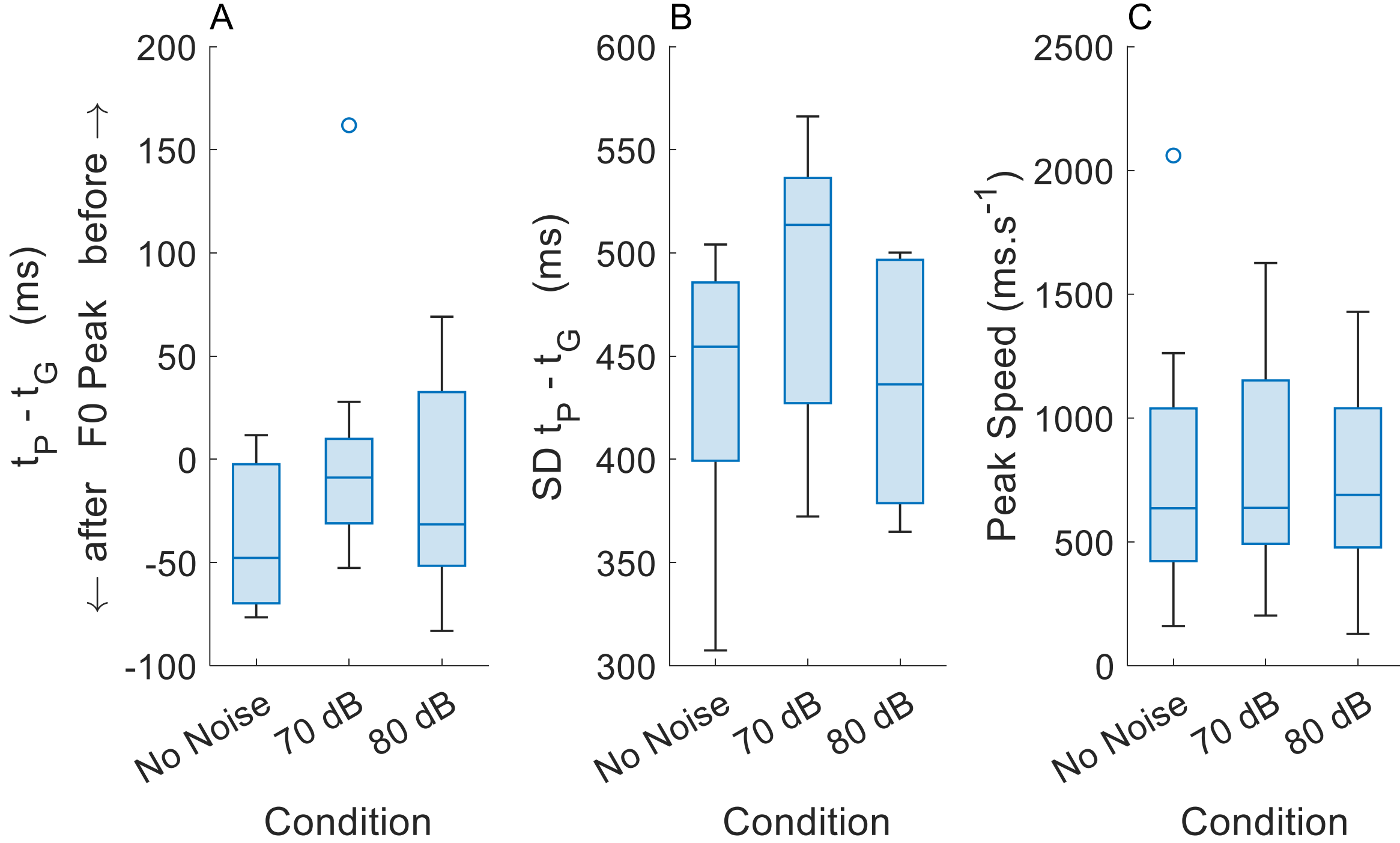

*Figure 8 The box-plot analysis of speech-hand synchrony of talker's beat gestures for different background noise levels (x-axis). (A) Across-subject mean value of statistics* $d$. *Values above 0 indicate that the gesture precedes the F0 peak, values below 0 indicate that the dominant gesture lags the F0 peak (B) Across-subject value of the standard deviation of statistics* $d$. *(C) Peak speed of the dominant movement.*

The coupling of speech with hand gestures was tested using the analysis of the distribution of statistics $d$. Figure 8A shows the across-subject distribution of mean values of statistics $d$, panel B shows the across-subject distribution of the standard deviation of statistics $d$, and panel C shows the peak values of the gestural movements that were used in the creation of the statistics $d$. Three box plots are displayed for three conditions of the background noise level in each of the panels.

The hand gesture peaks and speech pitch peaks (synchrony) did not show statistically significant shift in the means across noise levels, but we observed a statistically significant change in standard deviations (Table 12). The No Noise and 80 dB conditions were statistically significant from 70 dB condition (Bonferroni corrected pair-wise comparison showed $p<0.01$). Peak gesture speed was unaffected by noise.

| Source | F value | p value | p-adjusted | Significance | Partial $\omega^2$ |
|---|---|---|---|---|---|
| **Condition** | 4.99 | 0.0232 | 0.0256 | * | 0.3629 |

Table 12 ANOVA of the standard deviation of the $d$ statistics. Significance level: * $p<0.05$

# Discussion

The results demonstrate that when background noise during conversation increases, people produce more complex hand gestures in both roles: as speakers and listeners. Unsurprisingly, speakers use more gestures than listeners, and this holds across all observed hand gesture categories. The increase of hand gesturing during speaking is complemented by a reduction of idle postures, but the hand posture is not sensitive to noise conditions. The head movements adapt to rising noise levels: up–down head movements, shaking, and nodding increase in response to higher noise levels, aligning with the observed increase in hand-gesture complexity. Listeners' head movements are mainly characterized by nods; other types are rarely observed. The trunk movements are somewhat sensitive to increased noise levels. A decrease in leaning is complemented by an increase in leaning forward as the noise level increases. In terms of leg movements, walking in place is typically observed during about 10% of listening and speaking time, but no specific pattern emerged with respect to noise manipulations or mode of analysis (speaker vs. listener).

The speech sound level increases by 5 dB for every 10 dB increase in background noise. Further, we observed a 0.7dB –1.4 dB statistically significant increase in speech production related to the presence of hand movements, which was not affected by background noise (Table 10: interaction of condition and gesture type was not statistically significant). The head movements accompanying speech did not lead to an overall increase in speech production.

The analysis of hand–speech synchrony reveals a statistically significant modulation of the standard deviation of the $d$ statistic, but no shift. The increase is evident in the 70 dB condition with respect to the No Noise condition or the 80 dB condition. Statistic $d$ expresses a signed time difference between the peak in the utterance and the peak in hand movement velocity of the accompanying gesture, and an increase of standard deviation indicates broadening of the distribution that is obtained across many utterances in each condition. We interpret this as a decrease in hand-speech synchronization. The peak velocity of the selected gestures was, however, unaffected by the noise conditions, so qualitatively the gestures did not change in response to acoustic conditions.

The observed data support our first hypothesis that increasing the background noise level affects the frequency of gestures. Our observations of increased frequency of complex gestures with increased background noise level align well with the increased frequency and complexity observed in the communication task (Trujillo et al., 2021). In the previous experiments, the participant 'Producer' performed a one-way explanation of action verbs to the 'Addressee', with a one-way screen between them limited visual feedback. In our experiment, both participants interacted in a usual conversational setting without any obstructions. Since our data qualitatively and quantitatively align across the studies, it is possible that the increased hand gesturing reflects compensation for degraded auditory feedback and supports one-way production. Our data on hand postures show a modest decrease in idle time (Static drop posture), which may reflect increased hand movements. Our analysis did not specifically address hesitations in speech and hand movements (Betz et al., 2023).

In the first hypothesis, we also expect that head movements may support speech production in noise and be used more frequently by listeners to increase backchanneling or optimize SNR. The speakers generally made more head gestures: shaking, tilting, up/down, which were mainly present during speaking and almost completely absent during listening. Also, we see an increase in nodding – it most likely reflects the backchanneling when the addressee nods silently or nods with a confirmatory utterance – hence the distinction between the speaker may not reflect who holds the floor in the conversation. At the same time, head turning – which would reflect some SNR adaptations – is hardly present and does not show any sensitivity to changing background noise level. Together, hand movements during speaking may reflect speech prosody and visually support speech (lip movements are not analyzed). Nodding may reflect increased need for backchanneling in noisy situations. Head movements do not indicate SNR adaptations.

Within the first hypothesis, we also tested whether people change the frequency of trunk and leg movements in noisy conditions. Notably, people show increased frequency of leaning forward motions accompanied by a decrease in leaning side motions. This aligns with our expectations that people actively use transient movements towards each other (not only at the beginning when the noise conditions change

overall – which is coded as walking), therefore, the adaptation is active during the conversation and reacts to the ongoing pace, such as conversational breakdowns (not specifically coded). Here again, the roles of the listeners and speakers may not reflect the floor holder, since some of these movements were reflecting backchanneling with accompanying speech. Leg movements do not show any specific adaptation; from the current analysis, they do not play a significant role in adaptive behaviors.

The second research question tested whether hand and possibly head gesturing support speech production biomechanically in noisy situations (Pouw et al., 2020) or whether a compensation mechanism operates, with increased gesturing compensating for increased speech levels (Trujillo et al., 2021). Our data support the former because we observe a small (0.7dB –1.4 dB) but statistically significant increase in speech production related to hand movements across all conditions, with approximately equal amounts, even in No Noise. Plus, we found only limited support for the interaction of noise conditions and hand gesture types (the analysis also included hand postures). The support for this interaction comes from the data in the contractive posture condition. However, this posture included some movement; therefore, the static drop posture is a better reference for no movement, and there we could clearly see a separation from the other gesture types by 1.4 dB in all noise conditions. The available literature does not directly evaluate the dB SPL contribution of gesturing due to biomechanical coupling but mentions significant effects on F0 or intensity (Pouw et al., 2020, 2021, 2025). Cravotta et al., (2019) evaluated the effect of encouraged gesturing vs. no instructions during story retelling, finding a 0.8 dB difference in mean intensity. Most of the variance in the speech production data was due to the noise condition. Here, we could see a big difference between the noise conditions reflecting the standard Lombard effect. Our observations correspond to 0.5 dB of speech production increase for 1 dB increase, which corresponds well with the values cited previously (Gardner, 1971; Hodgson et al., 2007; Weisser et al., 2021).

Our third research question tested whether background noise in conversations influences speech production in terms of hand-speech synchrony and hand peak velocity. Our data show a decrease in synchronization (indicated by broadening of hand-speech delay distribution) but no influence on peak hand velocity. Previous research on hand-speech synchrony involved a color-matching task and manipulated the task difficulty by changing the placement of the icons on the tablet screen (De Jonge-Hoekstra et al., 2021). The participants had to name their actions during the task. Since the methodology of unstructured conversation widely differs from the color matching task, it is difficult to make direct connections or comparisons with that study. Perhaps one connection is that communication in noisy settings is subjectively more demanding than conversation in noise, but it is also possible that the two tasks were governed by very different cognitive mechanisms – but that needs to be clarified in future studies. A particular observation in our data is that the standard deviations increased only for the 70 dB noise condition, but not as much for the 80 dB noise condition. This may reflect motivation under conditions with increasing noise level that is often described by the inverted U-curve (Pichora-Fuller et al., 2016). Here, the engagement could have been highest in the mid-level condition and dropped for the higher levels, but the exact nature of such an effect on hand-speech synchrony has not been previously reported and should also be investigated in future studies.

A limitation of the present study is a relatively small sample (8 participants); extrapolations to a population level should be done with caution. Another limitation is the novelty of the proposed labeling system. The methodology is developed based on our own observations and available literature, but future iterations may refine some definitions and categories. The definitions served as a guide for the annotators, but more rigorous guidance may need to be developed in the future. Alternatively, the manual annotation may be replaced by an automated system (Białek et al., 2026), while the categories of typical movements could be extracted directly from observed data. In particular, the head and trunk movements are performed at different magnitudes, which made the cut-off for the No Movement category volatile; on the other hand, it did not affect the large movements, which are easily classified, therefore it is unlikely to substantially affect the present results.

Collectively, these results paint a picture in which the multimodality of the Lombard effect is reflected mainly in higher gesturing frequency of complex hand gestures and head movements, likely related to enhancing visual prosody in the dialogue. The hand gestures influence speech production sound level most likely at the biomechanical level, and our data do not show any additional feedback that would manifest as an increase or decrease in speech production level related to gesturing frequency. The observed changes in trunk movements may reflect a dynamic strategy to increase SNR during critical movements. The hand-speech synchronization decreased, as evidenced by a broader hand-speech delay distribution, which might be related to motivational factors in communication in noisy environments.

## Acknowledgements

Funded by the Deutsche Forschungsgemeinschaft (DFG, German Research Foundation) – Project ID 352015383 – SFB 1330 C5. Yang Jiao contributed to the development of the annotation system in his Master's thesis at the Technical University of Munich: Automatic Analysis of Body Movements and Gestures in Free Conversations. Cem Kaya helped with annotations and setting up the virtual environment. We thank all participants of the experiments.